# Pushing down the lateral dimension of single and coupled magnetic dots to the nanometric scale: characteristics and evolution of the spin-wave eigenmodes

Giovanni Carlotti

Dipartimento di Fisica e Geologia, University of Perugia, Via Pascoli, I-06123 Perugia, Italy

and  Centro S3, Istituto di Nanoscienze-CNR, I-41125 Modena, Italy

giovanni.carlotti@unipg.it

**ABSTRACT**

Planar magnetic nanoelements, either single- or multi-layered, are exploited in a variety of current or forthcoming spintronic and/or ICT devices, such as read heads, magnetic memory cells, spin-torque nano-oscillators, nanomagnetic logic circuits, magnonic crystals and artificial spin-ices. The lateral dimensions of the elemental magnetic components have been squeezed down during the last decade to a few tens of nanometers, but they are still an order of magnitude larger that the exchange correlation length of the constituent materials. This means that the spectrum of spin-wave eigenmodes, occurring in the GHz range, is relatively complex and cannot be described within a simple macrospin approximation. On the other hand, a detailed knowledge of the dynamical spectrum is needed to understand or to predict crucial characteristics of the devices.

With this focused review we aim at the analysis and the rationalization of the characteristics of the eigenmodes spectrum of magnetic nanodots, paying special attention to the following key points: (i) Consider and compare the case of in-plane and out of-plane orientation of the magnetization, as well as of single- and multi-layered dots, putting in evidence similarities and diversities, and proposing a unifying nomenclature and labelling scheme; (ii)  Underline the evolution of the spectrum when the lateral size of magnetic dots is squeezed down from hundreds to tens of nanometers, as in current devices, with emphasis given to the occurrence of soft modes and to the change of spatial localization of the fundamental mode for in-plane magnetized dots; (iii)  Extend the analysis from isolated elements to twins of dots, as well as to dense arrays of dipolarly interacting dots, showing how the discretized eigenmodes distinctive of the single element transform in finite-width frequency bands of spin waves propagating through the array.

**TABLE OF CONTENTS**

## I.  Introduction

Thanks to the advances in nanofabrication and miniaturization, magnetic devices based on single- or multi-layered constitutive elements, with lateral dimensions as small as a few tens of nanometers and with either shape or intrinsic magnetic anisotropy, are routinely produced for current or forthcoming applications.[1,2]  These include giant-magnetoresistance (GMR) sensors and read heads[3], magnetic memory cells[4,5] and bit-patterned hard-disks[6], spin-torque nano-oscillators[7] and nanomagnetic logic circuits,[8,9] artificial spin-ices[10,11] and magnonic crystals[12].  From the quasi-static point of view, the behavior of each of these small objects can be generally accounted for by assuming a unique magnetization vector (macro-spin approximation), but this simplified picture is not suitable to describe their high-frequency properties. In fact, these are characterized by the presence of several spin wave (SW) eigenmodes, occurring in the GHz frequency range, whose characteristics depend on the specific dot shape as well as on the direction and intensity of the external field. These excitations, also in combination with edge-effects and/or deviation from the ideal shape, can be at the origin of unexpected behavior that can significantly impact the proper design and realization of practical devices. For instance, spin wave eigenmodes may constitute a severe source of noise in read-heads, putting a limit to the rate of data writing or processing.[13] On the other hand, one can take advantage from specific SW eigenmodes to facilitate the switching of a magnetic dot, as in microwave-assisted magnetic recording[14,15] where the occurrence of soft modes can be relevant.[16] Finally, SWs can also be seen as a channel to dissipate power.[17,18,19]

Even if most of the previous studies of the magnetization dynamics in elliptical or rectangular dots dealt with magnetic dots with lateral dimensions above 200 nm,[20,21,22,23,24,] more recent studies have paid attention to the dynamics of dots as small as a few tens of nanometer, [25,26,27,28,29,30,31,32,33,34,35,36,37,38, 39,40,41] i.e. the targeted dimensions of many state-of-the-art devices. These dots are generally in the single domain state, although for sufficiently low values of the external magnetic fields and in presence of specific interactions with the substrate, it is also possible to achieve exotic configurations, such as vortex and bubble state.[42,43,44] The most powerful technique exploited to this aim is Brillouin light scattering (BLS), either conventional [45] or micro-focused,[46,47] although other powerful techniques have been developed, such as time-resolved magneto-optical Kerr effect (MOKE),[48] magnetic resonance force microscopy (MRFM)[49,50,51] thermally-activated electrical-noise measurements[41,52] and, more recently, scanning x-ray transmission microscopy (SXTM).[53,54] In parallel with the development of advanced experimental capability, also micromagnetic simulations and postprocessing data analysis were optimized to characterize not only the frequencies but also the spatial profile of eigenmodes resonating within the magnetic nanoelements.[20,55,56,57,58] Alternatively to micromagnetic simulations, other codes directly compute the susceptibility tensor in the frequency domain[42] or directly diagonalize the dynamical matrix.[59, 60] As stated above, these studies put in evidence that a detailed comprehension of the characteristics of the eigenmodes spectrum requires to go beyond the traditional macro-spin approximation, essentially because the lateral dimension of the dots are still at least one order of magnitude larger that the exchange length (4-5 nm in common ferromagnetic materials). Moreover, the internal field is not uniform, as it would be for an ellipsoidal body where the ferromagnetic resonance would be described by the simple Kittel formula,[61] so that localized modes appear in specific regions where the internal field is strongly non-uniform. Additional complexity appears when the magnetic dot consists of different layers, that may be coupled by either exchange or dipolar fields, so that one may observe the appearance of eigenmodes whose magnetization precession is in-phase (acoustic) or out-of-phase (optical) in the two layers. [52, 62,63,64,] Similarly, the degeneracy of discrete modes is lifted when the dots are arranged in twins[65,66] small clusters[67] or in one-[12] or two-dimensional[68,69,70] dense arrays of dots, constituting artificial magnonic crystals. In these systems, the anisotropic and spatially nonuniform magnetodipolar coupling among the dots leads to a collective dynamics that cannot be described in terms of spin wave modes of individual elements. In fact, in analogy to the well-known description of a systems of coupled oscillators in modern solid state physics, the collective dynamics is characterized by normal modes occurring at new frequencies, different from those of the individual (uncoupled) oscillators. For example, for a system of two magnetic layers or a twin of adjacent dots, one finds the splitting of the spin wave eigenmode in the above mentioned in-phase or out-of-phase modes, while for a large array of interacting dots collective (propagating) normal modes develop, leading to the formation of permitted and forbidden frequency bands.

In this focused review, we center our attention on the magnetization dynamics of magnetic dots with sub-200 nm lateral size in the single domain state, considering and comparing the case of in-plane and out-of-plane orientation of the magnetization and proposing a unifying nomenclature and labelling scheme. The characteristic features that occur in the eigenmodes spectrum when the lateral size of the considered dots is squeezed down from hundreds to tens of nanometers will be described, with emphasis given to the role of exchange-energy contribution, to the occurrence of soft modes and to the change of spatial localization of the main modes, in particular of the fundamental one. Special attention will be also paid to the effect of the chiral Dzyaloshinskii-Moriya interaction (DMI) that can be induced by an adjacent layer of heavy metal atoms with large spin-orbit coupling. [71,72,73,74,75] Then we will consider the modification of the spectrum when twins of interacting dots are considered, as a first step to extend our analysis to the case of either one- or two-dimensional dense arrays of dots, known as artificial magnonic crystals. Finally, we will generalize the above analysis to the case of multilayered structures, as spin valves or magnetoresistive sensors that involve at least two or three magnetic layers, separated by non-magnetic interlayers.

## II.  Eigenmodes of an isolated magnetic nanodot

Fig. 1 presents a schematic diagram illustrating the spatial characteristics and the labelling scheme for the magnetic eigenmodes of a circular magnetic nano-disk with diameter in the range of a few tens of nanometers, magnetized either out-of-plane (Fig. 1a) or in-plane (Fig 1b). It can be seen that in both cases the modes can be labelled using a couple of indices that are related to the presence of nodal lines.  In order to qualitatively comprehend the physical characteristics of these eigenmodes and their spatial character, let us recall that the equilibrium configuration of the magnetization can be found, in the micromagnetic approach, by minimizing the total free energy functional $E_{tot}$, that results from the sum of different terms, such as the Zeeman energy due to the external field $\boldsymbol{H}_0$, the exchange energy, the anisotropy energy and the magnetostatic energy. In addition, in presence of a heavy-metal under- or over-layer, one may have also an energy contribution coming from the above-mentioned DMI interaction.

As for the dynamics of the system, it is governed by the well known Landau-Lifshitz-Gilbert equation (LLG):

$$\frac{\partial \boldsymbol{m}}{\partial t} = -\gamma \left( \boldsymbol{m} \times \boldsymbol{H}_{eff} \right) + a \left( \boldsymbol{m} \times \frac{\partial \boldsymbol{m}}{\partial t} \right) \qquad (1)$$

where $\boldsymbol{m}(\boldsymbol{r}, t) = \frac{\boldsymbol{M}}{M_s}$ is the unit vector along the local magnetization, $\gamma$ is the gyromagnetic ratio, $\alpha$ is the Gilbert damping constant. $\boldsymbol{H}_{eff}$ is the effective field acting on the precessing spins, that can be derived from the energy functional according to:

$$\boldsymbol{H}_{eff} = -\frac{1}{\mu_0} \frac{\partial E_{tot}}{\partial \boldsymbol{m}} \qquad (2)$$

Therefore, it consists of different contributions, reflecting the energy terms recalled above:

$$\boldsymbol{H}_{eff} = \boldsymbol{H}_0 + \boldsymbol{H}_{exch} + \boldsymbol{H}_{ani} + \boldsymbol{H}_{ms} + \boldsymbol{H}_{DMI} =$$

$$= \boldsymbol{H}_0 + \frac{2A}{\mu_0 M_s}\nabla^2\boldsymbol{m} + \frac{2K_u}{\mu_0 M_s}m_z\,\boldsymbol{e}_z + \boldsymbol{H}_{ms} + \frac{2D}{\mu_0 M_s}\left[\frac{\partial m_z}{\partial x}\boldsymbol{e}_x + \frac{\partial m_z}{\partial y}\boldsymbol{e}_y - \left(\frac{\partial m_x}{\partial x} + \frac{\partial m_y}{\partial y}\right)\boldsymbol{e}_z\right] \tag{3}$$

where $\boldsymbol{H}_0$ is the external applied field, $A$ is the exchange stiffness constant, $K_u$ is the uniaxial perpendicular anisotropy constant and $D$ the DMI constant. $\boldsymbol{e}_x$, $\boldsymbol{e}_y$ and $\boldsymbol{e}_z$ are the unitary vectors of the reference frame, chosen such that the $z$ axis is perpendicular to the sample plane. The magnetostatic dipolar field $\boldsymbol{H}_{ms}$ would be uniform in the case of an ellipsoidal body, only, so that for planar dots it is usually calculated numerically, accounting for its non-uniformity and non-locality.

The most general method to solve the LLG equation above are based on proper numerical techniques where it is spatially discretized using finite differences or finite elements methods. As a result, a discretized version of the effective field is obtained and the corresponding system of differential equations are solved within suitable time-stepping schemes.

## A. Out-of-plane magnetization

Let us start from the analysis of the eigenmodes existing in a perpendicularly magnetized circular dot that represents an interesting model system for the free layer of perpendicular spin torque oscillators[76] or for the next generation of bit-patterned perpendicular magnetic media that will be used in forthcoming hard disks.[77] In a recent study,[56] we have exploited micromagnetic simulations, using a customized version of the commercial software Micromagus[78] to analyze the eigenmodes excited by either a pulse of magnetic field or a polarized injected current, in a circular free layer of Permalloy with thickness d=5 nm and two different values of the radius, namely R=50 nm and R=150 nm. The sample was perpendicularly magnetized thanks to the application of a relatively strong bias magnetic field (H_z=10 kOe) directed along the z axis, as shown in the top panel of Fig.2. At a given time, a weak pulse of field (duration 10 ps and constant intensity h_y=100 Oe), directed along the x axis, was applied, so to excite spin waves eigenmodes in our range of interest, i.e. up to and above 15 GHz. The ringing down of the magnetization of each discretized cell was then recorded for a period of 50 ns (with a sampling time of 10 ps) and the Fourier transform was calculated. Therefore, following the Micromagnetic Spectral Mapping Technique (MSMT),[20] both the eigenmode spectrum, averaged over the whole set of discretized cells, and the spatial profile of each eigenmode could be achieved. As illustrated in Fig. 1a, given the circular symmetry of the sample, the modes can be labelled according to a scheme based on a couple of indices *(r,l)*, where the radial number *r* counts the circular nodal lines and the azimuthal number *l* accounts for a phase shift of *l*×2π along a circular line. The plus and minus signs of *l* indicate the sense of the SW propagation in the counter-clockwise and clockwise directions, respectively. It can be seen in Fig 2a and 2b that the main mode is the "fundamental" (0,0) mode, characterized by a rather uniform spatial profile, i.e the absence of any nodal plane, with maximum amplitude in the dot center. Its frequency is appreciably larger in the case of R=50 nm (f=7.12 GHz) if compared to the case of R=150 (f=4.49 GHz), because of both the reduction of the perpendicular demagnetizing field and the exchange-energy contribution that are sizeable in the smaller dot.

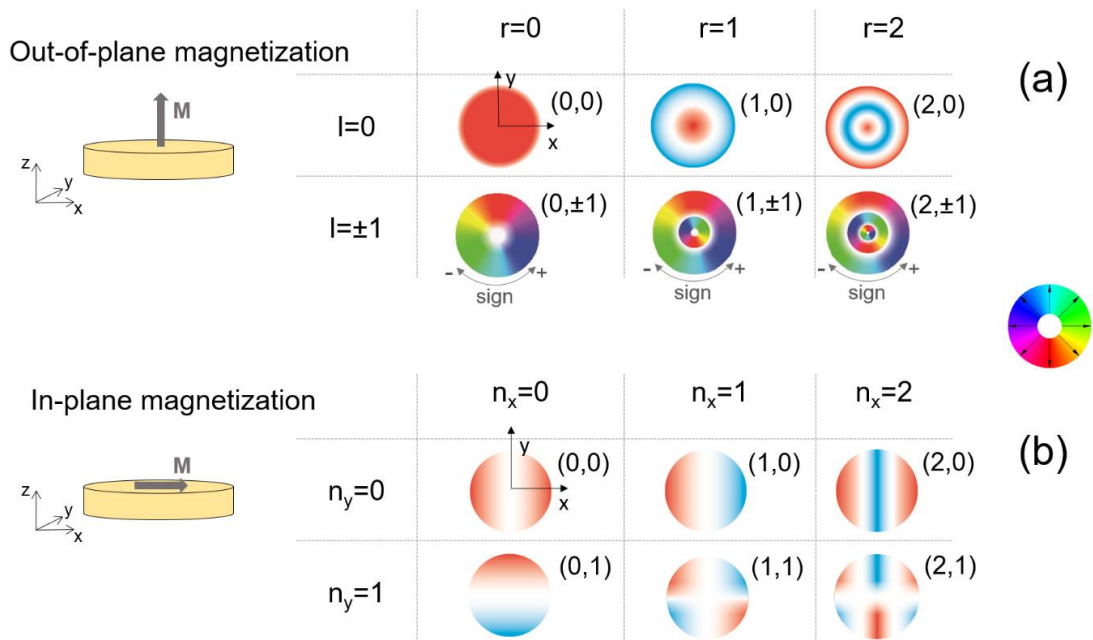

**Fig. 1** Schematic diagram showing the spatial characteristics and the labelling scheme for the magnetic eigenmodes of circular magnetic nano-disks with diameter in the range of a few tens of nanometers, magnetized either out-of-plane (a) or in-plane (b). In our coding scheme, the hue indicates the phase of the dynamical magnetization, according to the enclosed color wheel, while the brightness represents its amplitude. The nodal lines are marked in white.

Higher-order radial modes, labelled (1,0) and (2,0) are also visible at larger frequency for R=150 nm. Note that the frequency of the modes increases quite rapidly with increasing the number of nodal lines, essentially because of the increase of the exchange-field contribution $H_{exch}$ in eq. 3. By repeating the above "virtual experiment" and adding to the static field $H_0$ also the Oersted field generated by the injection of a current (J=9×$10^{10}$ A/m$^2$) flowing perpendicularly through the free layer, additional modes appear in the simulated spectra, as a consequence of the symmetry breaking induced by the Oersted field of the current. In particular, one can see the appearance, in addition to the above radial modes, of azimuthal (orthoradial) modes, labelled (0,1), (1,1) and (2,1), characterized by the presence of a 2π azimuthal phase-change (Fig. 2c and 2d). Looking at the dependence of the spectrum on the dot size, it emerges that the number of modes present within a limited frequency range increases with the dot radius. The fundamental (0,0) mode is always the most intense for excitation with a uniform pulse only, while it can be overcome by the (0,1) mode in presence of the Oe field that is characterized by a cylindrical symmetry.

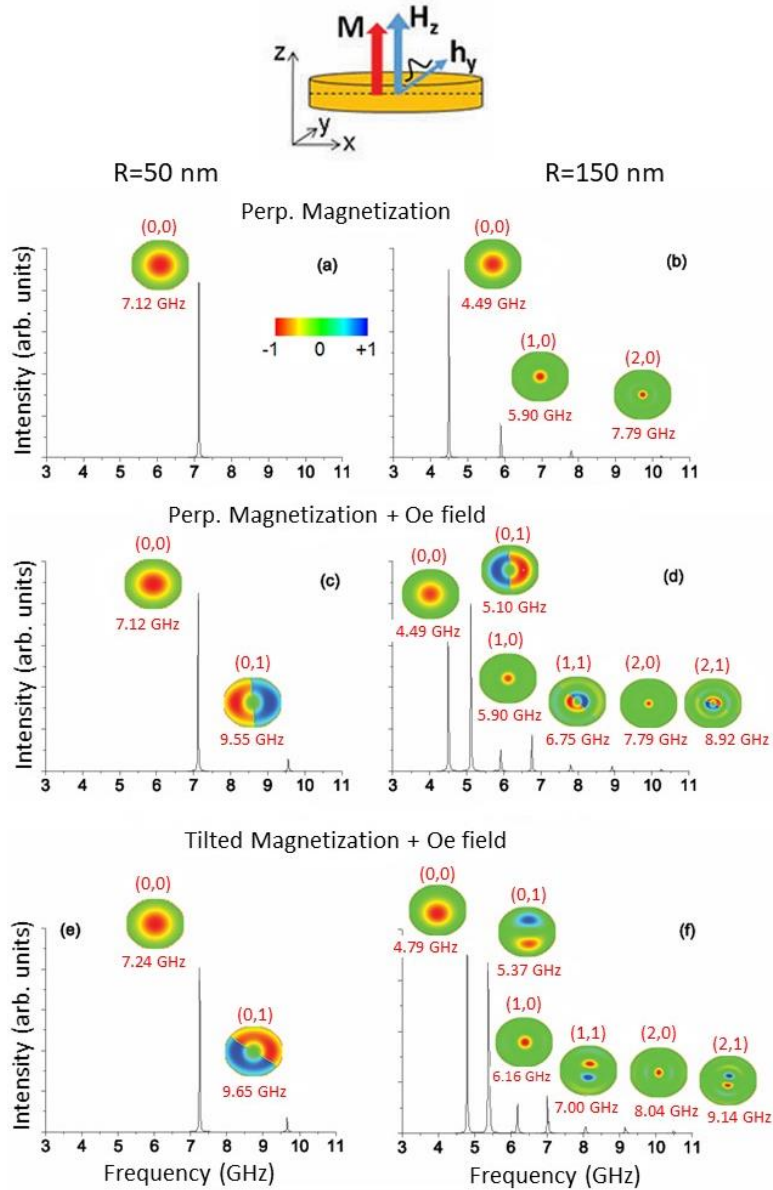

**Fig. 2** Simulated power spectra obtained after excitation of the dot by a uniform pulse of field ($h_x$=100 Oe, 10 ps long), while applying a constant external field $H_z$=10 kOe (i.e. $\mu_0 H$= 1 T) along the perpendicular direction (a-d). For the bottom panel (e-f) the external field H is tilted by 1° away from the z axis. The colour panels represent the spatial profile of the dynamical magnetization, expressed as the product of the modulus of the dynamical magnetization by the sign of its phase. Panels (c) to (f) refer to simulations where the presence of the Oersted field generated by a perpendicular current with density J=9×10¹⁰ A/m² is considered, in addition to the external static field. Left (a,c,e) and right (b, d, f) panels refer to a dot radius R=50 nm and 150 nm, respectively. Reproduced with permission from J. Phys D: Appl. Phys. **48,** 415001 (2015) and integrated with further panels. Copyright (2015) by the Institute of Physics Publishing (IoP). [56]

The fundamental mode frequency decreases with increasing the dot radius, attaining the ferromagnetic resonance frequency of the extended film for R larger than about 1 micron. Also the other modes decrease their frequencies with increasing R, due to the reduction of the internal field. It is also interesting to consider what happens if the external field H would not be applied exactly along the z axis. As seen in Fig. 2e and 2f, it is sufficient a tilt of 1° off the z-axis to induce remarkable variations of the spatial profiles of the modes, at least for the larger dot considered here. In fact, the symmetry breaking between the x and y directions induced by the tilt of the applied field, manifests itself in the loss of circular character of azimuthal modes (0,1), (1,1) and (2,1) in the dot with R=150 nm. Finally, let us note that if we would have considered elliptical dots, rather than circular, one would have observed qualitatively similar spatial profiles, with an elliptical (rather than circular, contour of the red and blue areas, but the labelling scheme and the characteristics of the radial and azimuthal modes would have remained qualitatively the same discussed above.

A second recent study was concerned with the eigenmodes of perpendicularly-magnetized circular dots of FeB with radius R=60 nm.[79] The authors observed thermally excited SW modes using electrical-noise measurements[41] and successfully compared the experimental results with both micromagnetic simulations and a simple perturbation theory. The observed modes were classified as eigenmodes with radial and azimuthal nodal lines, in agreement with the scheme proposed in Fig. 1a. Interestingly, they were able to provide evidence for the splitting of azimuthal modes (+ or − sign of $l$ in Fig. 1a), caused by the dynamic dipolar coupling that was fully taken into account.

A similar splitting of the azimuthal eigenmodes, caused this time by the presence of the above mentioned interfacial DMI interaction , was found theoretically in either circular or square magnetic dots of lateral size L=100 nm.[80] As illustrated in Fig. 3, even if the authors did not use the labelling scheme proposed in this review, it is clear that their Mode 1 corresponds to the (0,0) fundamental mode of our Fig. 1a, while the other modes are either radial (Mode 4, corresponding to (1,0)) or azimuthal ones (Mode 2 and 3, corresponding to (0,±1) and (0,±2), respectively). Interestingly, it was found that the presence of a sizeable DMI constant D leads to frequency splitting of the azimuthal eigenmodes, while the others are only slightly affected by the DMI. The magnitude of the splitting appears to increase with the value of the D and to be larger for the circular magnetic dots if compared to the squared ones. The frequency splitting is clearly associated with lifting in the degeneracy of eigenmodes $l=\pm l$. On the other hand, modes with a strong radial character, such as Modes 1 and 4, experience only a slight decrease in their frequency with increasing D and little change in their spatial profile. Similar features are also seen in the square dots, but the distinction between "radial" and "azimuthal" modes is not as sharp. In any case, it is evident that a careful analysis of the above-discussed frequency-splitting of some of the eigenmodes offers a tool to experimentally estimate the DMI constant in layered systems relevant for spintronics.

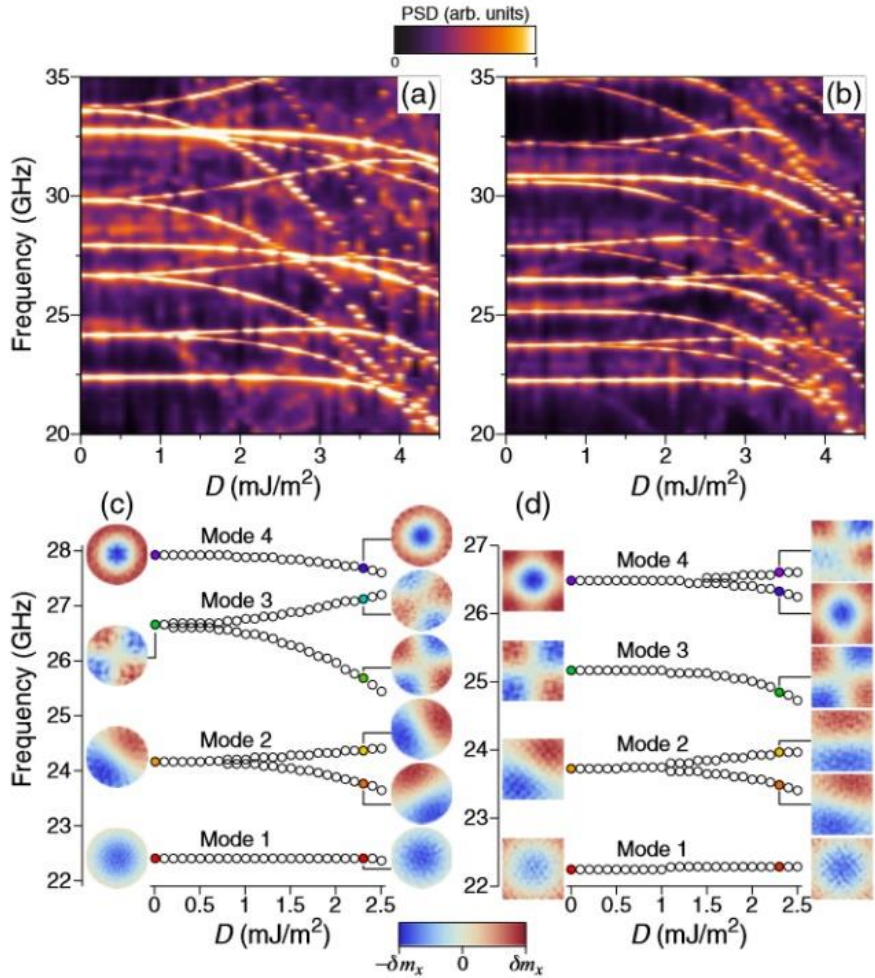

**Fig. 3** Map of the eigenmode power spectral density as a function of the interface Dzyaloshinskii-Moriya constant D for (a) 100-nm-diameter circular dots and (b) 100-nm-wide square dots. Selected profiles of the four lowest modes for different strengths of the DMI for the (c) circular and (d) square dots. Reproduced with permission from Phys. Rev. **B 89**, 224408 (2014). Copyright (2014) by the American Physical Society. [80]

## B. In-plane magnetized dots: effect of the inhomogeneous internal field

If we now consider magnetic nanodots with in-plane magnetization, then the internal field becomes strongly non-uniform so for absent or weak external fields one observes the formation of an inhomogeneous magnetization configuration, with flower, *C-* or *S*-states.[81, 21] In second place, even in presence of a saturating external field applied in-plane, the spatial profiles of the SW eigenmodes can be rather complex and evolve significantly with the lateral dimension of the dots. Similar to the case of the perpendicularly magnetized dots, each mode can be labeled with two integer indices ($n_x, n_y$) whose values correspond to the number of nodal lines. In this case however, as illustrated in Fig. 1b, the nodal lines are not radial and azimuthal, as in the case of perpendicular magnetization, but parallel or perpendicular to the direction of the magnetization (x axis). To illustrate the complexity the eigenmode spectrum in a realistic sample, let us start from the micromagnetic results relevant to an elliptical dot having dimensions $100 \times 60 \times 5$ nm$^3$, that were published in Ref. 57. The dot was discretized in cells with size of $1 \times 1 \times 5$ nm$^3$ (the small lateral size was chosen in order to better mimic the dot curvature) and its dynamical properties simulated using the Micromagus software.[82] Instead of using a pulse of field to excite the eigenmodes, as in the previous section IIA, here a stochastic magnetic field, uncorrelated in both space and time, was used to model thermal fluctuations. In such a way, one can obtain the whole set of modes existing in the nanodot (rather than a subset of modes compatible with the spatial symmetry of the pulsed field), but at the expense of a rather long simulation time and of somehow noisy spectra and profiles, as shown in Fig. 4. The trajectory of the magnetization of each discretized cell at T=300 K was recorded for 200 ns, at a constant value of the external field applied along the major axis. These data were used to determine the frequency and the spatial profile of the eigenmodes using a local Fourier-transform analysis. The power spectrum $P_{i,j}(f)$, of the magnetization was calculated for each discretization cell located at ($i,j$). Then, the average power spectrum was calculated as the sum of the power spectra of each single cell. The two-dimensional spatial distribution of the dynamical magnetization for each eigenmode was then determined by the MSMT technique anticipated in the previous paragraph.[20] The profiles are presented in Fig. 4 as bi-dimensional plots of the dynamical magnetization amplitude multiplied by the sign of its phase, together with the spectra calculated at room temperature for different values of an external field H applied along the major axis of the ellipse. It can be seen that there are several modes whose position in frequency evolves with the applied field intensity. Note that all modes display nodal planes, except the (0,0) one (notice that the presence of the white central region in mode (0,0) of Fig.1b does not correspond to a nodal line, because the phase of the dynamical magnetization does not change sign across this boundary). This would be the only mode present in the spectrum if one calculates the Fourier transform of the time averaged magnetization, as shown in the lowest spectrum of Fig. 4. Therefore, from the experimental point of view, the (0,0) mode is the "fundamental" mode, that would be the only one detected in either a ferromagnetic resonance, or a Brillouin light scattering experiment at small incidence angle of light. Magnetostatic effects appear in the fact that the (0,0) mode amplitude is much larger at the edges than at the dot center. In fact, the magnetostatic (demagnetizing) field $H_{ms}$ is relatively large close to the dot borders orthogonal to the static magnetization, because of the appearance of free magnetic poles. As a consequence, the effective field $H_{eff}$ is much lower close to the dot borders where the dynamic amplitude becomes larger than elsewhere. This characteristic localization may affect the frequency and the linewidth of this mode since it is highly sensitive to the structural quality/imperfections of the dot borders imperfections,[50] as well as to size and shape variations in large arrays of dots.[31]

Similarly to the case of the perpendicularly magnetized dots, also in the present case the dot frequency increases quite rapidly with the mode indices, essentially because the introduction of nodal lines causes a substantial increase of the exchange-field contribution $H_{exch}$. In Fig. 5 the evolution of the eigenmodes frequency with the external field is shown. The frequency decreases with decreasing external field, reflecting the reduction of the effective field felt by the precessing spins.

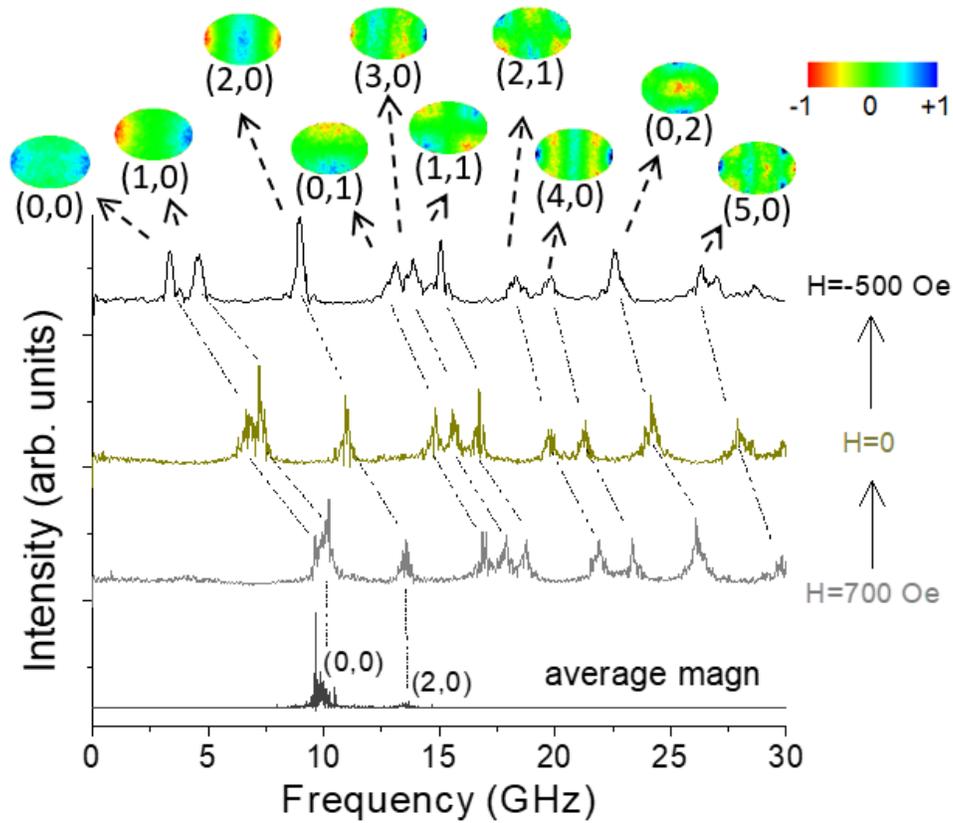

**Fig. 4** Simulated eigenmodes spectra of the elliptical dot for different values of the external field H, applied along the major axis of the ellipse. The three upper spectra are calculated as the average from the spectra of the discretized cells, while bottom spectrum is relative to the average magnetization of the dot. The top insets illustrate the calculated spatial profiles of the relevant eigenmodes (the colors represent the amplitude of the dynamical magnetization multiplied by the sign of its phase). Reproduced from J. Appl. Phys. **115,** 17D119 (2014) with the permission of AIP Publishing and integrated with further panels. [57]

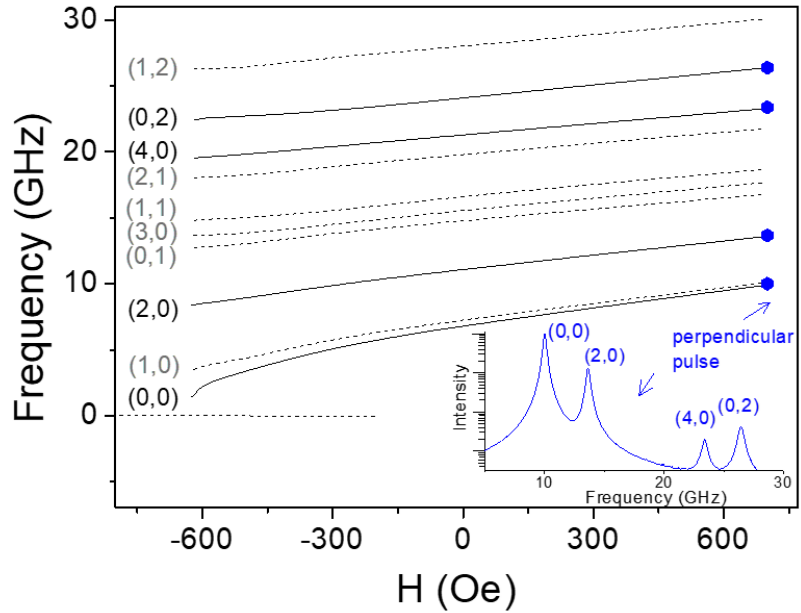

**Fig. 5** Evolution of the eigenmodes frequency with the intensity of the magnetic field, applied along the major axis of the elliptical dot, from +700 Oe to -625 Oe. The full circles correspond to the four modes that can be excited with an out-of-plane field pulse and the corresponding spectrum is reported as a bottom inset. Reproduced and adapted from J. Appl. Phys. **115,** 17D119 (2014) with the permission of AIP Publishing. [57]

Remarkably, when the field is reversed and its value approaches that of the coercive field (H=-630 Oe), the (0,0) mode softens, leading to the reversal of the dot magnetization, as will be discussed in section IIC. The bottom inset of Fig. 5 shows the spectrum that can be obtained exciting the system with a uniform perpendicular pulsed field: only spatially symmetric modes are excited.

It is also interesting to consider how the characteristics of the mode spectrum change when the lateral dimensions of the magnetic dot increase from the nanometric to the micrometric scale, according to the results of Ref. 39, that is extended here to consider dot with sub-100 nm dimension, with a thickness of 10 nm.[83] In this case, to reproduce the BLS measurements performed at almost normal incidence, in the micromagnetic simulations the modes were excited by a unifrom field pulse applied perpendicular to the film plane. This means that only spatially symmetric modes, (such as the previous mentioned (0,0) and (2,0)) modes could be efficiently excited. As seen in Fig. 6, when the dot length is below about 30 nm, the only mode that is present in the frequency range up to 20 GHz is the (0,0) mode. On increasing the dot length L above about 40 nm, one sees that also the (2,0) mode enters the relevant frequency window. For dot length above about L=100 nm, one observes that the lowest mode (0,0) tends to be strongly localized at the dot edges, since it gets trapped in the minima of the internal field close to the borders of the dot. Therefore it loses its "fundamental" character and can be labelled as symmetric edge mode (S-EM). Instead, the (2,0) mode, characterized by a large amplitude in the central region of the dot, for dot length above 100 nm becomes the mode with the largest average dynamical magnetization and its frequency approaches that of the uniform ferromagnetic frequency of the extended film. Therefore, as seen in the insets of Figs. 6 and 7, for L above about 100 nm this mode can be labelled as "fundamental" (F) or "center mode" and it is by far the most intense of the spectrum, when it is excited by a spatially unifrom external field (as it happens in microwave assisted switching, for instance). Instead, for L below 90 nm, the most intense peak is the (0,0) mode, with maximum amplitude at the dot edges. Remarkably, in the range of intermediate dot lengths between about 80 and 110 nm, the two modes are simoultaneously excited with similar intensity, so that both of them are entitled to be defined as "fundamental" modes. This interchange of the "fundamental" character and the evolution of the modes frequencies with the length of the elliptical dot can be also followed in Fig. 7, where the region of coexistence of two "fundamental modes" is highlighted by the rectangular shaded area. Note that this simultaneous presence of two 'fundamental' modes, with comparable intensity can have important consequences in terms of coupling with an external uniform field or beating and mixing phenomena if the nonlinear regime would be achieved. In Fig. 7 we have also reported the frequency evolution of the antisymmetric (1,0) mode, that was not present in Fig. 6 since it can be excited only applying an antisymmetric field pulse to the dot. For L larger than 100 nm, this mode becomes an antisymmetric edge mode (AS-EM) and its frequency is very close to that of the S-EM. Instead, when L is reduced below about 90 nm, the presence of the nodal plane causes a large cost in terms of exchange energy and it becomes the usual (1,0) mode of smal dots, with a frequency that increases monotonously with reducing the dot dimension.

Note that in the case of dots with lateral size L above 200-300 nm,[35, 84] in addition to the center mode (F) and the edge modes (EM) reported in Figs. 6 and 7, there are several other modes resulting from the evolution of the $(n_x, n_y)$ modes described above. They are dominated by the magnetostatic energy for large dot size, while the exchange-energy contribution becomes less important, and assume the labelling of $n_x$-BA (backward) modes (dipolar modes with $n_x$ nodal lines perpendicular to H) and $n_y$-DE (Damon–Eshbach) modes (with $n_y$ nodal lines parallel to the direction of H). Remarkably, in relatively large dots, where the contribution of the exchange field $\boldsymbol{H}_{exch}$ to the eigenmode frequency is much lower that of the magnetostatic dipolar field contribution $\boldsymbol{H}_{ms}$, the BA modes appear at frequencies below that of the F mode. This is consistent with the fact that magnetostatic backward volume waves (magnetostatic surface waves, i.e Damon-Eshbach waves) in thin films have frequencies below (above) the ferromagnetic resonance.[85]

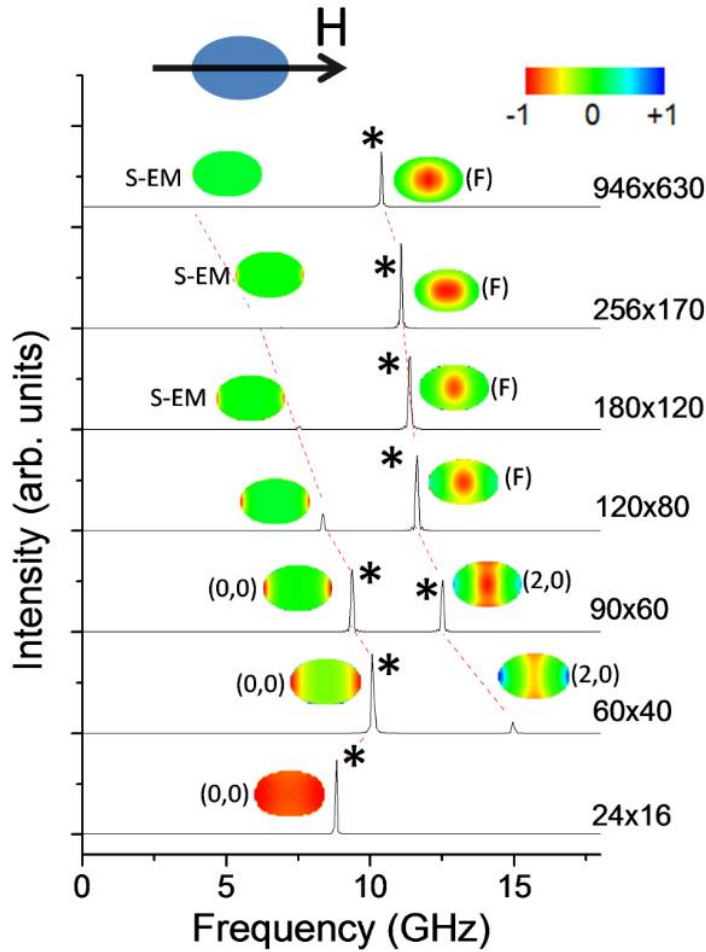

**Fig. 6** Simulated power spectra obtained after excitation of the dot by a uniform perpendicular pulse of field (10 Oe high and 10 ps long), while applying a constant external field H=1 kOe along either the easy in-plane direction of the elliptical dots. For each spectrum, the dimensions of the long and short axis of the dot (expressed in nm) are reported. The red lines are guides for the eye, while the asterisk connotes the mode with the "fundamental character" in each spectrum. The color panels represent the spatial profile of the dynamical magnetization of selected eigenmodes, expressed as the product of the modulus of the magnetization by the sign of the phase. Please note that antisymmetric modes, such as the (1,0) mode, are not present in these spectra because they are not excited by the spatially uniform pulse of field used here. Reproduced with permission from J. Phys. D: Appl. Phys. **47,** 265001 (2014) and extended to lower dimensions of the dots with further panels. Copyright (2014) by the Institute of Physics Publishing (IoP). [39]

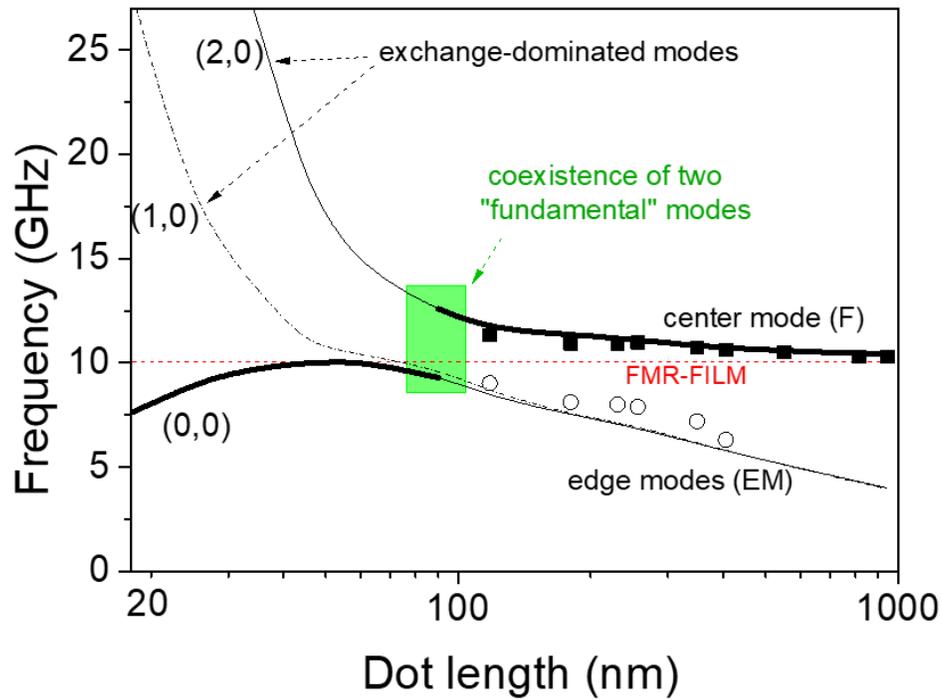

**Fig. 7** Evolution of the calculated (lines) and measured (points) frequencies of the three main modes present in a small (in-plane magnetized) elliptical dot, as a function of the dot length (major axis of the ellipse) for the external magnetic field H=1 kOe, applied along the easy in-plane direction. It is seen that the (0,0) and (1,0) modes of sub-100 nm dots become the symmetric and antisymmetric edge-modes of large dots, while the (2,0) mode evolves in the center mode of large dots. The filled (bold) aspect of the dots (curves) indicate the most intense peak in experimental (simulated) spectra, respectively. Within the highlighted rectangular area there is the coexistence and the crossing of the "fundamental" character of the modes, as explained in the text. Reproduced with permission from J. Phys. D: Appl. Phys. **47,** 265001 (2014), adapted, integrated with further data and extended to lower dimensions of the dots. Copyright (2014) by the Institute of Physics Publishing (IoP). [39]

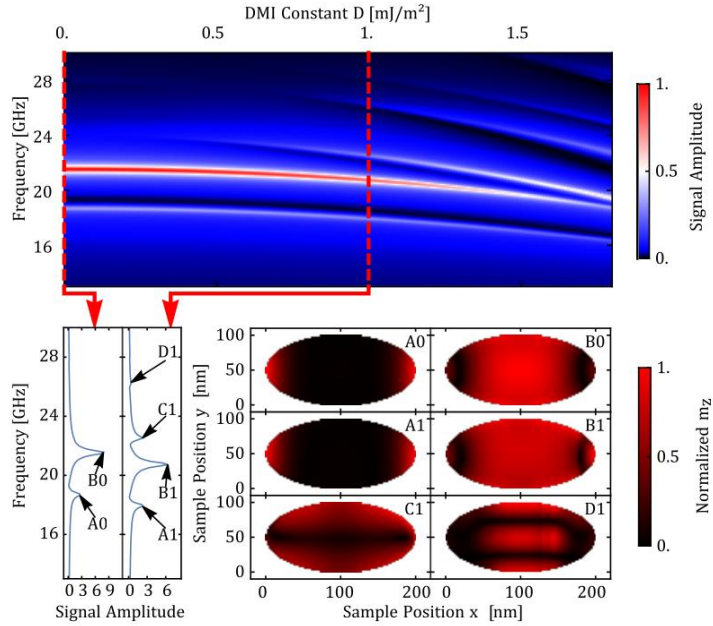

Fig. 8 Micromagnetic calculations showing how mode intensities and profiles depend upon interfacial DMI for a 200 nm x 100 nm elliptical ferromagnet. The upper plot summarizes the frequencies and intensities of modes as a function of the DMI strength. Shown in the lower left are mode spectra calculated by taking the Fourier transform in time after excitation by a spatially uniform pulse of field. The corresponding mode intensity profiles are shown on the right. Reproduced with permission from Phys. Rev. **B 99**, 214429 (2019). Copyright (2019) by the American Physical Society. [86]

Let us conclude this section dedicated to the characteristics of SW eigenmdes in in-plane magnetized dots, citing a very recent paper that has been just published concerning the influence of DMI interaction on the modes spectra.[86] The authors found that, as a consequence of the nonreciprocity of SW propagation along +k and -k, in a confined geometry states with well-defined nodes which are inherently phase modulated such that space-inversion symmetry of the mode profile is lost. Therefore, additional spectral features become visible in ferromagnetic resonance studies of microelements with DMI, as illustrated in Fig. 8 for the case of a 200 nm × 100 nm elliptical dot. It is evident that modes labelled A0 and B0 correspond to modes (0,0) and (2,0) of previous Figs. 6 and 7, also named "edge mode" and "center mode" for increasing dot size. Moreover, it is found that for D < 1 mJ/m² , the shift in frequency is moderate and additional modes faintly appear, while for values D > 1 mJ/m² the shift of the eigenfrequencies becomes substantial and high-order modes appear in the spectra,  as a clear fingerprint of DMI.

### C. Soft modes and microwave-assisted switching

The above analysis about the spectrum of the eigenmodes of a magnetic dot and its evolution with the applied field can be important in order to achieve efficient switching. In particular, involvement of eigenmodes in the magnetization reversal process can be particularly relevant when one has the occurrence of soft modes as well as when one exploits microwave-assisted switching.[14,15]

Soft magnetic modes may appear when the system undergoes a transition between different magnetic configurations. "Softening" here refers to the progressive decrease of a mode frequency as the external field gradually approaches the critical transition field. For example, it is seen in Fig. 5 that in proximity of the coercive field of a 100 nm × 60 nm elliptical dot, the (0,0) mode becomes soft: the reversal starts at the dots edges (where the (0,0) mode has maximum amplitude) and rapidly involves the whole surface of the dot. In larger dots, instead, the situation is more complicated: usually, either an EM or a BA mode becomes soft and triggers the magnetization reversal that proceeds through a sequence of complex intermediate states (including vortices). A series of joint experimental and micromagnetic studies have been carried on to analyze the connection between soft modes and either continuous or discontinuous transitions of the magnetization in nanodots[16,87,88,89] or nanorings[84,90]. The experimental investigations have been carried out by the BLS technique, following the frequency evolution of spin wave modes on the applied magnetic field. These works revealed the dynamic origin of the reversal process and identified the spin mode responsible for the onset of the instabilities that lead to reversal. At the critical field, the system becomes unstable with respect to this excitation, due to the vanishing of the magnetic restoring forces: hence, the previous magnetic state is no longer an equilibrium one and a new final configuration is pursued through a sequence of non-equilibrium states and nonlinear processes. Actually, the magnitude of the space profile of the specific soft mode is directly connected with the area of instability (which occurs where the precession amplitude is larger at the transition), while the mode symmetry provides information about the path followed by the system to accomplish the first steps of the magnetic transition.

The relationship between eigenmodes and reversal process is even more important if one considers recent achievements in the field of microwave assisted switching, i.e. the excitation of the precessional motion of the magnetization by a radio frequency field that is used to achieve a substantial reduction of the coercive field necessary for the switching.[15] In this respect, a detailed knowledge of the relationship between the characteristics of the eigenmode spectrum and the dot geometry, anticipated in the previous two sections IIA and IIB, may become extremely relevant. For instance, Fig. 9, taken from Ref. 91, shows a color contour plot of the coercive field of an in-plane magnetized elliptical dot of Py, with lateral size of 65 nm × 71 nm and thickness 10 nm, as a function of both the frequency and the amplitude of an applied microwave signal. It can be seen that at 2.15 GHz, the coercive field was reduced by 26% with a rf field amplitude of only 0.4 mT.

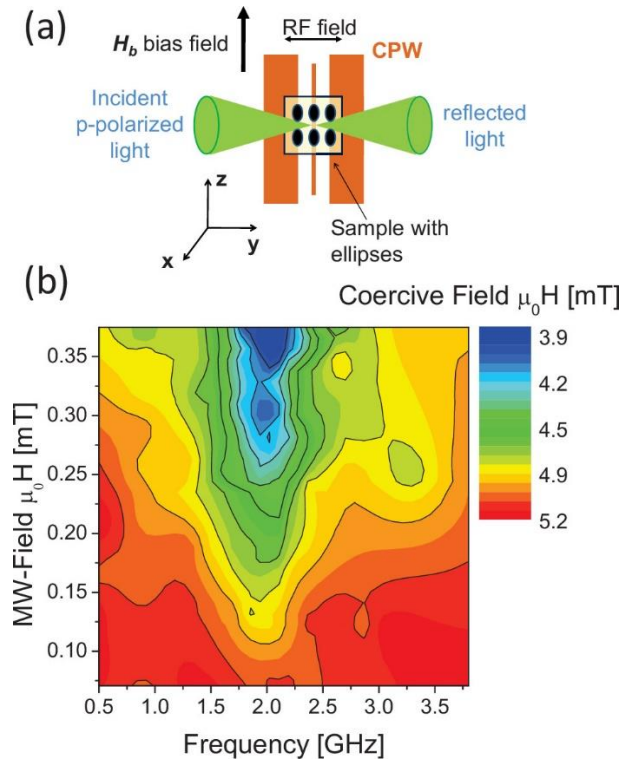

**Fig. 9** (a) Setup for measuring the hysteresis curves with an applied rf-field. (b) Contour plot of the coercive field as functions of rf field and frequency. The coercive field is strongly reduced at microwave frequencies in the region around 2 GHz and decreases as the microwave field amplitude increases. Reproduced from Appl. Phys. Lett. **95** 062506 (2009) with the permission of AIP Publishing. [91]

Similar reductions of the coercive field were also observed in a variety of soft [14,92,93,94,95,96,97,98] and hard[99,100,101,102,103,104] magnetic materials, magnetized either in-plane or out-of-plane, such as thin films and nanodots. In the latter case, knowledge of the spatial characteristics of the eigenmodes is important since, as we have seen in Sects. IIA and IIB, spatially non-uniform magnetization precession is largely present in nanodots. For instance, it was shown in Ref. 15 that for a perpendicularly magnetized circular magnetic dot, similar to those discussed in Sect. II.A, not only the (0,0) mode, but also higher order radial modes could be excited by the microwave field, resulting in a notable increase of the efficiency of switching. In practice, this is already exploited in current devices for microwave-assisted magnetic recording, where the rf field is generated by a spin-torque oscillator integrated in the writing head of the hard disk.[105,106,107,108]

To conclude this section, we want to mention recent interesting results relative to microwave-assisted switching in arrays of Py dots acting as nanomagnetic logic elements that were studied experimentally by coplanar broadband ferromagnetic resonance (FMR) and numerically by micromagnetic simulations.[109] It was found that edge-mode excitation by a microwave field could be a feasible way to address a specific element in the cluster. In particular, a driver-input magnet pair demonstrated a significant reduction of the switching field by the excitation of the edge mode located at the uncoupled end of the driver.

## III. Clusters and arrays of closely spaced dots

### A. Twins of dots

In this section we move to the analysis of coupled dots and we start from considering what happens to the modes of rectangular nanodots, with lateral dimensions around 100 nm, when they are not isolated, as in the previous sections, but are placed in twins, as shown in Fig. 10a, taken from Ref. 66. In that particular study, Brillouin scattering spectra were measured and compared to the simulations for isolated dots; then, the effect of interdot dipolar coupling on the SW eigenmodes of twins of coupled dots, placed either head-to-tail or side-by-side, was analyzed. Three kinds of dots were studied, having a fixed value of the shorter side d=60 nm and three different values of the longer side, i.e. D=90, 120 and 150 nm, respectively. For each dot dimension, three arrays were prepared, as shown in Fig. 10a for the case of the dots with D= 90 nm. A first array where the dots are isolated (or stand-alone, SA), being the separation among adjacent dots chosen to insure negligible interaction; a second array where the dots are arranged in twins placed head-to-tail (HT); a third array where the dots are arranged in twins placed side by side (SS). In the two latter cases, the separation of the dots within each couple is about 20 nm.

According to what was discussed in the previous paragraph, the dots considered here are in the range of lateral dimensions where the simultaneous presence of two "fundamental modes" of comparable intensity should occur. In fact, this is confirmed by both the BLS experiments shown in Fig. 10b and the micromagnetic simulations of the power spectra of the average magnetization excited by a uniform perpendicular pulse, reported in Fig. 10c. The two main peaks seen in the spectra correspond to the mode (2,0), localized in the center, and the mode (0,0) localized at the edges of each dot. Remarkably, when the dots are put HT (SS), the effect of dipolar interaction causes a blue-shift (red-shift) of the center mode frequency by a fraction of a GHz with respect to the SA configuration.

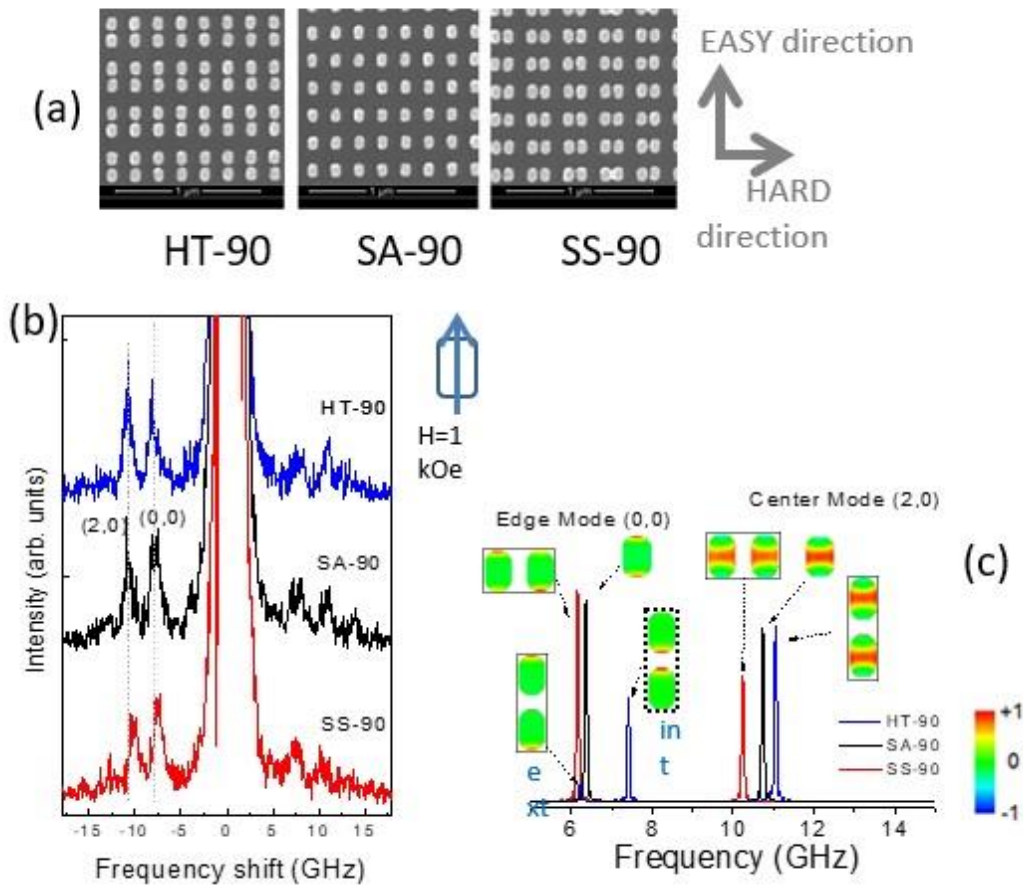

**Fig. 10** (a) Scanning electron microscopy images of the samples, where the labels HT, SA and SS refer to head-to-tail, stand-alone and side-by-side dots, respectively. Measured (b) and simulated (c) spectra for an external magnetic field H=1 kOe applied along the easy direction. Reproduced and adapted from J. Appl. Phys. **117**, 17A316 (2015) [66] with the permission of AIP Publishing.

Finally, in the simulations it appears that for the HT arrangement, the edge-mode splits into two modes: one at roughly the same frequency as in the SA or SS arrangements, localized at the external edges and another one upshifted by about 1 GHz, localized at the internal edges, where the internal field is larger due to interdot coupling. Details about the evolution of the measured frequencies with the dot length D, for the three sets of samples, as well as the discussion of the characteristics of the eigenmodes when the field is applied along the hard direction, can be found in Ref. 66.

A similar study concerning twins of rectangular magnetic nanodots ($100 \times 50 \times 10$ nm$^3$), where the splitting of the eigenmodes due to the inter-element magneto-dipole interaction, was also performed in Ref. 110. In this study, the excitation was performed by both a uniform (symmetric) field pulse (as in the study reported above) and an antisymmetric field pulse, so to excite not only the modes (0,0) and (2,0) (labelled in the original paper Acoustic Bulk (AB) and Acoustic Edge (AE)), but also the antisymmetric (1,0) and (3.0) modes (labelled in the original paper Optical Bulk (OB) and Optical Edge (OE)). In particular, it was shown that for the edge modes the interaction between the edges of neighbouring elements can exceed that between the edges of the same element, leading to softening of the mode profile. Moreover, the difference in frequency of the symmetric (acoustic) and antisymmetric (optical) modes was taken as a measure of the strength of the interaction between different elements.

## B. One- and two-dimensional arrays of dots: magnonic crystals

The effects of dipolar interdot coupling have been also widely investigated in the case of dense arrays of magnetic elements, constituting what is known as a magnonic crystal. This, in fact, can be formed starting from uncoupled magnetic elements and making them coupled by magneto-dipolar interaction, so that the standing waves of individual elements can interact via dipolar coupling. As a consequence, the degeneracy of the discrete eigenmode frequencies of different elements is removed and magnonic bands of dispersive, Bloch-type, excitations are formed, with a frequency amplitude that depends on the particular type of standing spin mode. To this respect, it is evident that the magnitude of these magnonic effects depends not only on the geometry of the array of dots (lateral size, thickness and separation between adjacent dots), but also on material parameters. In particular, stronger dipolar coupling and broader magnonic bands are expected for materials with larger saturation magnetization. A recent review article[12] collected the most relevant contributions this field that are however concerned with chains[111, 112] or 2-d arrays[69, 113,114,115,116] of relatively large magnetic dots (hundreds of nm). Here however, as stated in the introduction, we want to focus on results relative to sub-200 nm nanodots, so we recall first of all the interesting pioneering study [68] performed by time-resolved scanning Kerr microscopy to image collective spin wave modes within a 2D array of 80 nm × 40 nm magnetic nanoelements arranged in a square matrix with inter-element separation of 20 nm. The most intense peak in the simulated spectra of the single nanodot was the (0,0) fundamental mode, as expected. However, if one considers a finite array of 3×3 dots, different peaks appear in the simulated spectrum. In particular, for a bias field of 197 Oe, as seen in Fig. 11, one finds three peaks: at the frequency of the highest peak (4.8 GHz) all elements precess in phase. Instead, the peak at the frequency of 4.5 GHz (5.4 GHz) corresponds to the center row (column) precessing out of phase with respect to the rest of the elements, while the amplitude of precession is increased in the center row (column).

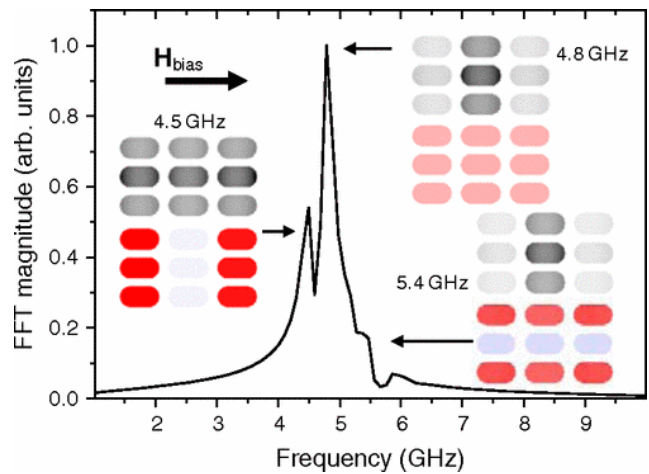

**Fig. 11** Simulated spectrum of a the 3 x 3 array of nanodots, for a bias field of 197 Oe. The insets show the magnitude (top) and phase (bottom) for the modes at 4.5 GHz, 4.8 GHz, and 5.4 GHz. Reproduced with permission from Phys. Rev. Lett., **104**, 027201 (2010). Copyright (2010) by the American Physical Society. [68]

These findings allowed the authors to ascribe the splitting observed in the simulations to collective nonuniform precessional modes of the 3×3 array. The mode at 4.8 GHz can be classified as quasi-uniform, while the modes at 4.5 GHz and 5.4 GHz are the collective backward-volume-like and Damon-Eshbach–like modes, respectively. Magneto-optical Kerr effect measurement confirmed the splitting of the fundamental mode peak, proving that the collective modes extend through the entire measured array. This means that spin waves are confined within the array as if it was a single element made of a continuous material, so that such arrays appear as tailored magnonic metamaterials for spin waves with a wavelength much greater than the period of the array.

Another interesting dense array of small nanodots (240 nm × 80 nm), that has been recently investigated, was an artificial spin-ice (ASI) system,[117] where couples of dots were fashioned into square ASI-like geometry. Using Brillouin light scattering, the frequencies of excitations were measured as a function of the magnetic field, showing that the frequencies of spin waves localized at element edges evolve non-monotonically with magnetic fields and soften at certain critical fields. From measurements of such critical fields, the authors were able to extract information on the magnetization reversal of individual islands within the array.

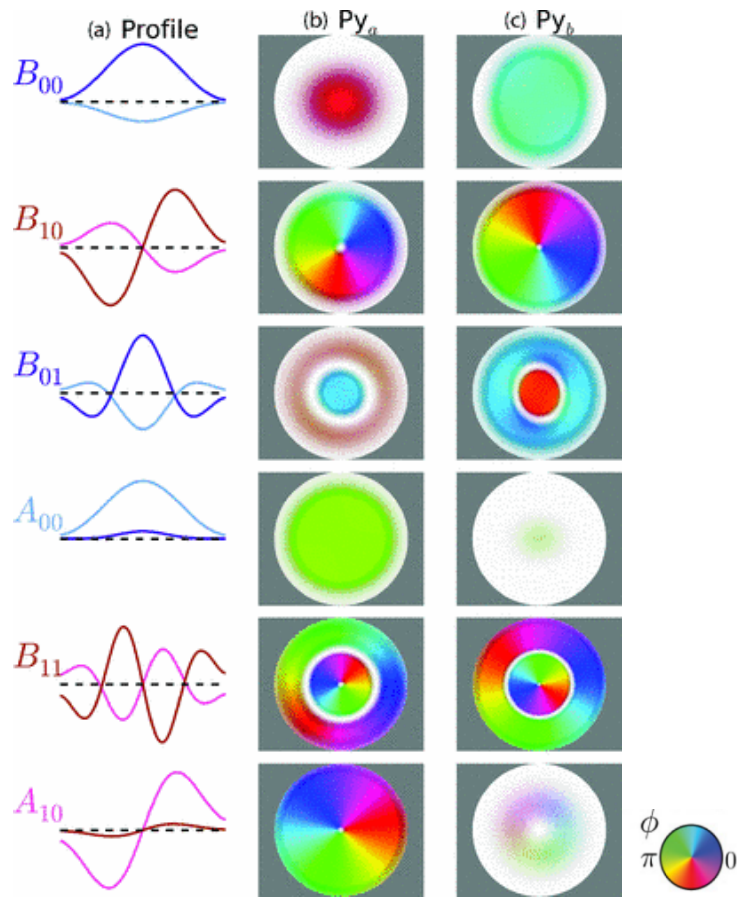

**Fig. 12** Simulated precession patterns for a number of relevant eigenmodes existing in the perpendicularly magnetized pillar consisting of two permalloy layers, 4 nm and 15 nm thick, separated by a Cu interlayer 10 nm thick. Column (a) shows the precession profiles across the thin (light color) and thick (dark color) layers. Columns (b) and (c) show the dynamics in the thin $Py_a$ and thick $Py_b$ layers, respectively. In our coding scheme, the hue indicates the phase φ of the dynamical magnetization, while the brightness its amplitude. The nodal lines are marked in white. Reproduced from Phys. Rev. B **84**, 224423 (2011), under Creative Commons License. [49]

## IV. Multilayered dots

In some applications and devices, such as spin torque oscillators or magnetoresistive read heads, the magnetic elements consist of two or more magnetic layers separated by a non-magnetic spacer. Therefore, in this last Section we want to consider the case of a multilayered magnetic element. As in the previous case of single-layer nanodots, while several studies exist for relatively large multilayered magnetic dots,[63, 64,118] we want to focus here on two recent studies dedicated to either circular or rectangular mutilayered pillars with sub-200 nm lateral size.

### A. Perpendicularly magnetized circular pillars

The first study was concerned with perpendicularly-magnetized circular nano-pillars, with radius R=100 nm that were analyzed both experimentally, using mechanical ferromagnetic resonance, and theoretically, using analytical calculations and micromagnetic simulations.[49,119] The pillars consisted of two permalloy layers, labelled $Py_a$ and the thick $Py_b$, of different thickness $t_a = 4$ nm and $t_b = 15$ nm, sandwiching a 10-nm copper (Cu) spacer. Each of the isolated layers present eigenmodes similar to those introduced in Fig. 1a and 2, but the collective magnetization dynamics could be classified depending if the precession in the two layers occurs in phase (antibinding modes, or acoustic modes) or out-of-phase (binding modes or optical modes). Fig. 12 shows a gallery of calculated spatial profiles for a number of representative modes: the anti-binding ($A$) and binding ($B$) eigenfrequencies were calculated by an analytical model, labelling the modes as $A_{l,r}$ and $B_{l,r}$ were ($l,r$) refer to the number of azimuthal and radial nodal lines. Consistent results were also obtained by micromagnetic simulations. (Please note that this is the same labelling scheme anticipated in Fig. 1a for perpendicularly magnetized dots, but here the two indices are interchanged). One may notice in Fig. 12 that the binding (lower energy) mode $B$ always corresponds to a larger precession amplitude in the thicker layer, with the thin layer vibrating in antiphase, and vice versa for the anti-binding mode $A$. In particular, the fundamental binding mode $B_{0,0}$, has an amplitude of precession that is three times larger in the thick layer, while the amplitude of the fundamental anti-binding mode $A_{0,0}$, is eight times larger in the thin layer than in the thick one. The authors found a very good agreement between the measured and the simulated dynamics, as illustrated in Fig. 13. Noteworthy, they found that a linearly polarized, spatially uniform rf field, can only excite modes with zero azimuthal index (left panel of Fig. 13), while excitation of l=1 azimuthal modes can be achieved introducing an orthoradial (i.e. azimuthal) stimulus, for instance provided by an injected current (right panel of Fig. 13). This selective excitation of a subset of modes with the proper spatial symmetry is similar to what already observed for single-layered dots of Sect. IIA and II.B.

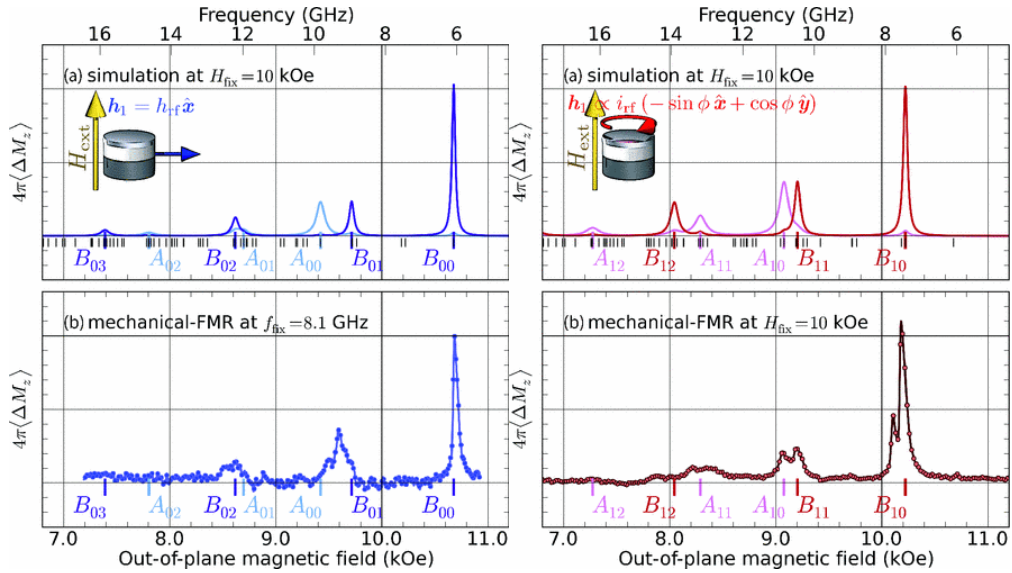

**Fig. 13** Left panel: (a) simulated spectral response to a uniform excitation field directed along the x-axis; (b) experimental spectrum measured by mechanical FMR exciting the nanopillar by a homogeneous rf magnetic field at 8.1 GHz. Right panel: (a) simulated spectral response to an orthoradial excitation field; (b) experimental spectrum measured by mechanical-force microscopy for an rf current excitation. Reproduced from Phys. Rev. B 84, 224423 (2011), under Creative Commons License. [49]

## B.   In plane magnetized bilayer: the read head

A couple of recent papers were devoted to a detailed analysis of the spin wave eigenmodes of a spin valve sensor for GMR read heads[120,121], with lateral dimensions as low as those exploited in state of the art devices (30 nm × 35 nm), whose detailed composition and magnetization orientation is sketched in Fig. 14a. In particular, one may notice that the free layer (FL) has a magnetization vector rotated by 90° with respect to the reference layer (RL), thanks to the application of an external field (that in real GMR sensors is provided by an integrated permanent magnet). Here, it was assumed that the FL is exposed to a bias field $H_b$= 600 Oe, uniform across the layer, that is sufficient to rotate the FL magnetization along the hard direction (*x*-axis), as seen in Fig. 14a. Both the eigenmodes spectrum, averaged over the whole set of discretized cells, and the spatial profile of each eigenmode within each layer could be obtained by micromagnetic simulations, applying a spatially uniform field pulse perpendicular to the plane of the layers. The scheme adopted for labelling the modes is based on that already anticipated in Sect. II, i.e. two indices ($n_x$, $n_y$) are used to indicate the number of nodal lines perpendicular or parallel to the direction of the static magnetization, respectively. Here the situation is more complex, because there are multiple layers and the static magnetization in the FL is orthogonal to that in the RL. Therefore, the authors introduced the tag 'FL' or 'RL' before the two indices ($n_x$, $n_y$) to identify which is the layer where the oscillation is preferentially localized. Using this scheme, it is seen in Fig. 14b that, despite the relatively small lateral dimensions of the considered sample, the spectrum is relatively rich and at least five well defined eigenmodes are seen in the frequency range up to 20 GHz. The solid and dashed curves in Fig. 14b refer to the spectra calculated averaging over the discretized cells of either the FL or the RL, respectively. It is clear that the dominant peak at about 2 GHz, labelled FL(0,0), corresponds to the fundamental mode of the FL, characterized by no nodal planes and by a nearly synchronous precession of the magnetization all over the layer surface. Remarkably, even if the lateral dimension of the pillar is only 30 nm, this is still several times larger than the exchange correlation length (about 4.5 nm) so that the inhomogeneity of the internal field induces a marked localization at the edges orthogonal to the static magnetization (longer edges). One can notice that this FL(0,0) mode has a remarkable amplitude also in the RL, because of the strong interaction between the two layers. However, in the RL the static magnetization is orthogonal to that of the FL, so this mode has maximum amplitude close to the two edges orthogonal to the *y*-axis (shorter edges). Moving to higher frequency, one finds at 7.9 GHz the fundamental eigenmode of the RL, labelled RL(0,0), localized at the shorter edges. This has a much smaller amplitude in the FL, because the thickness of the RL is less than a half compared to the FL, so it cannot drive efficiently the motion of the FL. At about 9.7 GHz and 15.3 GHz one can see the modes labelled FL(1,0) and FL(0,1) that are characterized by the presence of a nodal line either perpendicular or parallel to the direction of the RL magnetization, respectively. These two modes, whose amplitude is concentrated in the FL, can be considered as the antisymmetric counterpart of the FL(0,0) mode and are shifted to higher frequency mainly due to the exchange-related stiffness caused by the presence of the nodal planes. Finally, at about 18.4 GHz one finds the RL(1,0) mode, that is mainly localized in the RL and is the antisymmetric counterpart of the RL(0,0) mode. Its amplitude is also relatively intense in the FL, because it is localized in the same region of the FL(0,1) mode, that is not too far in frequency. In the original study[120] it was also analyzed the effect of the non-uniformity of the bias field produced by a realistic permanent magnet, showing that it reduces the inhomogeneity of the modes across the layers. Moreover, analysis of the influence of the discussed eigenmodes on the magnetoresistive readback signal revealed that when the amplitude of the precession of the magnetization exceeds a few degrees, nonlinear effects lead to the appearance of harmonics and peaks at the difference and sum of the main modes FL (0,0) and RL (0,0). The peak at the frequency difference contributes to a low-frequency tail, increasing the low-frequency noise that can somehow limit the optimum performance of read-head devices.[121]

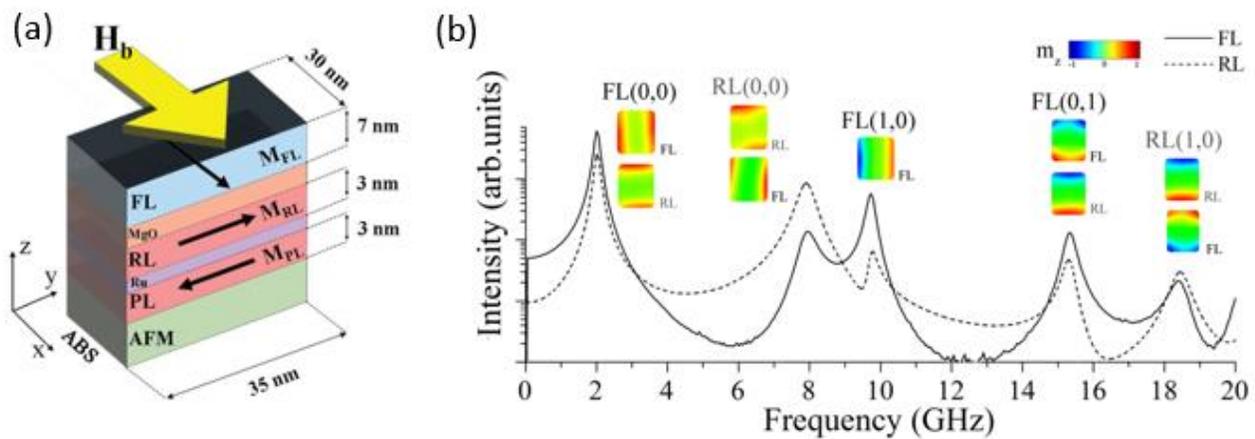

**Fig. 14** (a) Sketch of the simulated magnetic tunnel junction (MTJ) read head. The black arrows represent the orientation of the magnetization in each layer, while the yellow arrow is the direction of the bias field $H_b$ and ABS indicated the air bearing surface. (b) Simulated power spectra of the FL (solid line) and RL (dashed line) after the excitation pulse in presence a uniform bias field of $H_b$=600 Oe. The colored panels represent the spatial profile of the dynamical magnetization component perpendicular to the sample plane of the main eigenmodes, expressed as the product of the modulus of the magnetization by the sign of the phase. The label on the right side of each panel refers to the layer where the spatial profile is calculated. Reprinted with permission and adapted from J. Phys D: Appl. Phys. **50**, 455007 (2017). Copyright (2017) by the Institute of Physics Publishing (IoP). [120]

## V.        Summary and outlook

In conclusion, we have reviewed the characteristics of the eigenmode spectrum of in-plane and out-of-plane magnetized dots with lateral dimensions ranging from a few tens to a few hundreds of nanometers, that is typical of recent and forthcoming devices. We have shown that the eigenmode properties are influenced by different effective fields, derived from different energy terms (such as exchange, magnetostatic and anisotropy energy, as well as the chiral DMI contribution coming from a heavy metal underlayer) and one can distinguish different families of modes, labelled with two indices, according to the number of nodal lines. Particularly interesting is the case of in-plane magnetized dots: as the lateral size is reduced below about 200 nm, the exchange energy starts to play a significant role and one observes that the 'fundamental' character is progressively transferred from the center mode to a second mode having maximum amplitude at the edges of the dot. The amplitude of these two modes is comparable intensity in a range of lateral sizes around 100 nm that is typical of current devices, where the coexistence of two resonant frequencies may have relevant consequences in terms of coupling to external stimuli or noise characteristics.  More generally, a detailed knowledge of the eigenmodes spectrum and its evolution with the dot dimensions is important because its characteristics not only determine the small amplitude dynamics, but also influence nonlinear phenomena such as switching and large amplitude oscillations driven by an applied field or spin transfer torque.   When the dots are arranged in clusters or in dense arrays, as in magnonic metamaterials, the band of propagating collective excitations can be seen as resulting from the lifting of the degeneracy of eigenmodes of the single dot, thanks to dipolar coupling. Finally, for multilayered dots, currently exploited in spin-torque oscillators or magnetic memories, a correct identification of the eigenmodes can be achieved only considering that both magnetic layers take part in the field-induced or current-induced magnetization dynamics. On the other hand, one finds that the amplitude of coupled modes may be either equally distributed in the layers or concentrated in one of the layers, depending on the geometry and the strength of the coupling. This can affect the magnitude of the spin-transfer torque in a nanopillar or the magnetoresistive signal in a read-head sensor, since they depend on the coupled dynamics of all the magnetic layers.

**Acknowledgement**

Financial support from the EMPIR programme 17FUN08-TOPS, co-financed by the Participating States and from the European Union's Horizon 2020 research and innovation programme, is kindly acknowledged.

phenomenological damping constant, *Ms*=860 G for the saturation magnetization, *A*= 1. 3×10−6 erg/cm for the exchange stiffness.